\documentclass[a4paper, 12pt]{article}

\usepackage[colorlinks,allcolors=blue]{hyperref}
\usepackage{amssymb,amsmath,amsthm,mathrsfs}
\usepackage{graphicx, tikz, tikz-3dplot} 

\usepackage{wasysym}
\usepackage{color}

\theoremstyle{definition}
\newtheorem{thm}{Theorem}[section]

\newtheorem{rem}[thm]{Remark}

\def\R{\mathbb{R}}

\def\C{\mathbb{C}}
\def\N{\mathbb{N}}

\def\Z{\mathbb{Z}}

\def\Sh{\mathcal{S}}

\def\0{\bm{0}}

\def\H{\mathcal{H}}
\def\Ran{\mathrm{Ran}}

\def\Coker{\mathrm{Coker}}
\def\span{\mathrm{span}}
\def\e{\mathrm{e}}

\begin{document}
\title{Discrete time scattering and Wold's decomposition \\ in pictures}

\author{R. R. Firdaus${}^1$, S. Richard${}^{2}$
\footnote{Supported by JSPS Grant-in-Aid for scientific research C no 21K03292.}}

\date{\small}
\maketitle
\vspace{-1cm}

\begin{quote}
\begin{itemize}
\item[1] School of Sciences, Nagoya University, Furo-cho, Chikusa-ku, Nagoya, 464-8602, Japan
\item[2] Institute for Liberal Arts and Sciences, Nagoya University, Furo-cho, Chikusa-ku, Nagoya, 464-8601, Japan
\item[] E-mail:  rafi.rizf@gmail.com, richard@math.nagoya-u.ac.jp
\end{itemize}
\end{quote}

\begin{abstract}
Based on explicit computations, various concepts of discrete time scattering theory
are reviewed, discussed, and illustrated. 
The dynamics are taking place on a discrete half-space.
All operators are represented graphically.
The expressions obtained for the wave operators lead to an easily visualized interpretation
of Wold's decomposition, a seminal result of operator theory.
This work is clearly pedagogically oriented with the aim of providing explicit formulas and pictures
for usually unexplicit operators.
\end{abstract}

\section{Introduction}\label{sec:intro}

Scattering theory is a well-developed theory that provides asymptotic information about the evolution
of dynamical systems as time goes to plus or minus infinity. If the time parameter is continuous,
the evolution is mainly driven by unitary groups of the form $\{\e^{-itH}\}_{t\in \R}$ with $H$
a self-adjoint operator on a Hilbert space, while if the time parameter is discrete, 
one often considers unitary groups of the form $\{U^n\}_{n\in \Z}$ with $U$ a unitary operator. 

In this pedagogically oriented paper, we provide an example of a discrete time evolution system for which all computations can be performed
explicitly. As a consequence, various outcomes of scattering theory can be easily visualized, and several concepts
can be concretely discussed. The invariance of certain subspaces under the evolution,
the commutation relation of the scattering operator with the free evolution, 
the chain rule, all of these notions can be represented in pictures and checked graphically.
We also take the opportunity of getting explicit expressions to illustrate a
fundamental result of operator theory: Wold's decomposition. This result says that any isometry can be decomposed
into the sum of a unitary operator and a collection of shift operators. For all wave operators computed in this paper,
we directly exhibit this decomposition, without any further effort.

Let us describe more precisely the content of this paper. 
In Section \ref{sec:models}, we recall the main idea of discrete time scattering theory
and introduce the wave operators and the scattering operator. 
These operators are going to play a prominent role in the sequel.
We then immediately describe the scattering model we shall focus on. 
The framework consists of the lattice $\Z\times \N$ representing a discrete
two-dimensional half-space. Various unitary evolution groups
are defined on this space, by exhibiting the action of these operators 
on a natural basis of the corresponding Hilbert space $\ell^2(\Z\times \N)$.
With these concrete operators, explicit expressions and pictures are provided for the
wave operators and for the corresponding scattering operators.

In Section \ref{sec_outcomes}, we recall a few general properties of the wave operators
and exemplify them with our explicit models. Here, the intertwining relation
satisfied by the wave operators plays a crucial role. We discuss
the ranges of the wave operators and their cokernels. A special emphasis is put 
on the invariance of these subspaces with respect to the evolution groups.
Properties of the scattering operator are also discussed.
Even though the wave operators are highly non-trivial, the scattering operators
are quite simple. This fact is mainly due to its commutation relation 
with the free evolution group, as explained in this section.

With numerous explicit expressions for the wave operators available, 
we could not resist making the best use of them to illustrate a seminal result valid for all isometries: Wold's decomposition.
At no cost, this abstract result can be clearly visualized.
In Section \ref{sec_Wold} we recall this theorem and explain
how the previous expressions can be understood with this result in mind.
This section also supports our strong belief that the interaction between scattering theory
and abstract operator theory is not only interesting but also fruitful: the former provides
many interesting and meaningful examples while the latter organizes various results and
enlarges perspectives. 
In summary, our aim is to illustrate various concepts of scattering theory with examples and
pedagogical explanations. 

\section{Discrete time scattering theory and explicit expressions}\label{sec:models}

Let us start by recalling the main question and ideas of scattering theory, and provide some important definitions.
We concentrate on discrete time scattering theory. 
A related recent work based on discrete time evolution groups can be found in \cite{BMP}.

Given a unitary operator $U$ on a Hilbert space $\H$, and given an element $f\in \H$, can one find an auxiliary unitary operator $U_0$
and two elements $f_\pm\in \H$ such that the following asymptotic estimates hold:
$$
\lim_{n\to \pm \infty}\big\|U^n f - U_0^n f_{\pm}\big\| = 0~?
$$
Equivalently, these conditions also read
$$
\lim_{n\to \pm \infty} \big\|f-U^{-n}U_0^n f_{\pm}\big\|=0.
$$
Thus, if the limits exist in the sense mentioned above, and if we set\footnote{Note that
other notations are also used for $W_\pm(U,U_0)$ in the literature, and even the convention for
the $\pm$ sign depends on the authors.}
\begin{equation*}
W_\pm(U,U_0) f_\pm:=\lim_{n\to \pm \infty} U^{-n}U_0^n f_\pm,
\end{equation*}
then the following relations hold
$$
f=W_\pm(U,U_0) f_{\pm}.
$$
These equalities mean that the evolution $U^n f$ can be approximated by 
the evolution $U_0^n f_{\pm}$ for $n\to\pm \infty$ if and only if
$f$ belongs to the range of the so-called wave operators $W_\pm (U,U_0)$. 
Clearly, this approach is interesting only if $U_0$ is simpler than $U$.
In particular, $U_0$ is often assumed to be an operator with an absolutely continuous spectrum.

Let us also emphasize that not only the ranges of the wave operators are of interest, 
but their orthocomplement contain also interesting information. 
Indeed, any element of $\H$ orthogonal to the range of $W_\pm(U,U_0)$
can not be described asymptotically by the evolution defined by $U_0$.
It means that these elements either do not scatter, or scatter with an
asymptotic evolution which is not well approximated by $U_0$. Both situations can 
take place, as we shall see in the models introduced below.

\subsection{Initial model}\label{sec0}

In this subsection and the following two, we introduce the models\footnote{This initial model is inspired by \cite[Ex.~5.12]{Amrein}.} 
and provide the main results about scattering theory.
The configuration space for our models is the discrete half-space $\Z\times \N$, with $\N=\{0,1,2,3.\dots\}$.
Accordingly, the description takes place in the Hilbert space $\H:=\ell^2(\Z\times \N)$, 
endowed with the standard orthonormal basis 
\begin{equation}\label{eq_basis}
\{\delta_{x,j}\mid x\in \Z, j\in \N\}
\end{equation}
defined by $\delta_{x,j}(y,k)=1$ if $y=x$ and $k=j$, and $\delta_{x,j}(y,k)=0$ otherwise. 
For any $f,g\in \H$, the scalar product of these two elements is given by
$$
\langle f,g\rangle :=\sum_{x\in \Z} \sum_{j\in \N} \overline{f(x,j)} g(x,j)
$$
while the norm on this Hilbert space is defined by $\|f\|:=\sqrt{\langle f,f\rangle}$.

\begin{figure}[ht]
\centering
\begin{minipage}[b]{0.5\textwidth}
\centering
\includegraphics[width=0.7\textwidth]{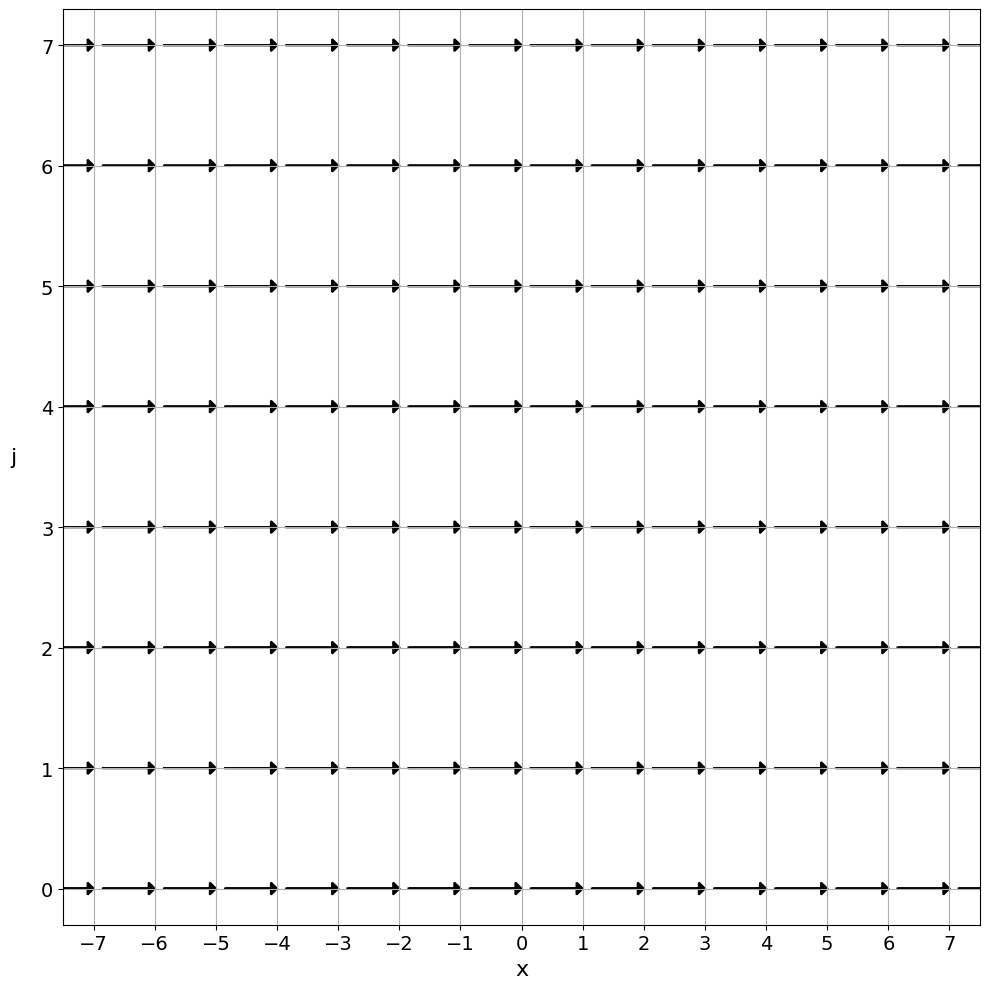}
\caption{Action of $U_0$}
\label{fig_U_0}
\end{minipage}%
\begin{minipage}[b]{0.5\textwidth}
\centering
\includegraphics[width=0.7\textwidth]{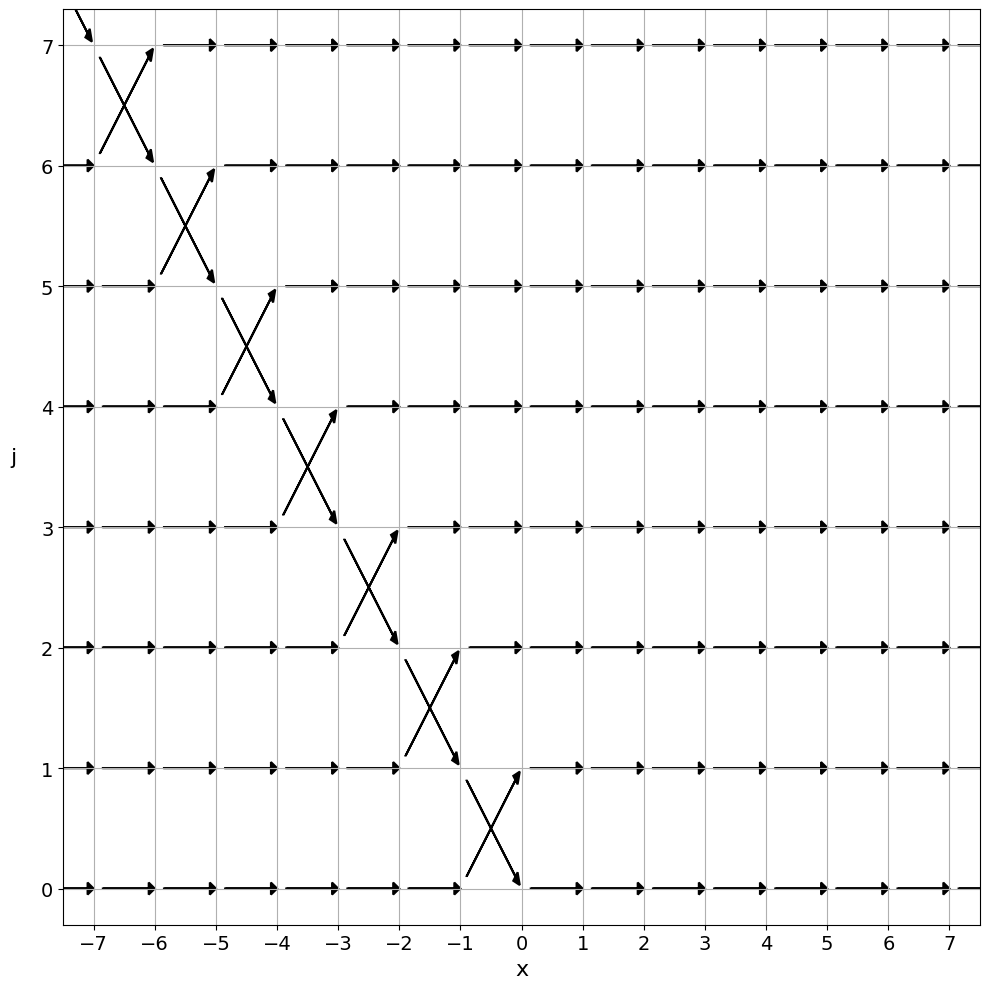}
\caption{Action of $U_1$}
\label{fig_U_1}
\end{minipage}%
\end{figure}

We consider two unitary operators $U_0$ and $U_1$ defined by their action on the standard basis, namely
$$
U_0 \delta_{x,j}:=\delta_{x+1,j}
$$
and 
$$
U_1\delta_{x,j}:=\begin{cases}
\delta_{x+1,j} & \hbox{ if }\  x\not \in \{-j-1, -j\}, \\
\delta_{x+1,j+1} & \hbox{ if }\  x =  -j-1, \\
\delta_{x+1,j-1} & \hbox{ if }\  x =  -j , n\neq 0,\\
\delta_{1,0} & \hbox{ if }\  x = 0, j= 0.
\end{cases}
$$
We refer to Figures \ref{fig_U_0} and \ref{fig_U_1} for the
actions of these operators.

\begin{rem}
Let us explain how to read the figures provided in this paper: each point on the lattice
can be associated to an element of the basis introduced in \eqref{eq_basis}. 
If $A$ corresponds to the operator represented in a figure, then an arrow 
starting at $(x,j)$ and ending at $(y,k)$ means $A\delta_{x,j}= \delta_{y,k}$.
A black dot at $(x,j)$ means that $A\delta_{x,j}=\delta_{x,j}$, or in other words the operator
$A$ acts as the identity on the element $\delta_{x,j}$ of the basis of $\H$.
\end{rem}

By using the scalar product on $\H$ and the relations
$\langle f,U_0^{-1}g\rangle = \langle U_0f,g\rangle$ and 
$\langle f,U_1^{-1}g\rangle = \langle U_1f,g\rangle$
one infers that
$U_0^{-1} \delta_{x,j}=\delta_{x-1,j}$ while 
$$
U_1^{-1}\delta_{x,j}=\begin{cases}
\delta_{x-1,j} & \hbox{ if }\  x\not \in \{-j, -j+1\}, \\
\delta_{x-1,j-1} & \hbox{ if }\  x =  -j+1, j\neq 0,\\
\delta_{x-1,j+1} & \hbox{ if }\  x =  -j,\\
\delta_{0,0} & \hbox{ if }\  x = 1, j= 0.
\end{cases}
$$

\begin{figure}[ht]
\centering
\begin{minipage}[b]{0.5\textwidth}
\centering
\includegraphics[width=0.7\textwidth]{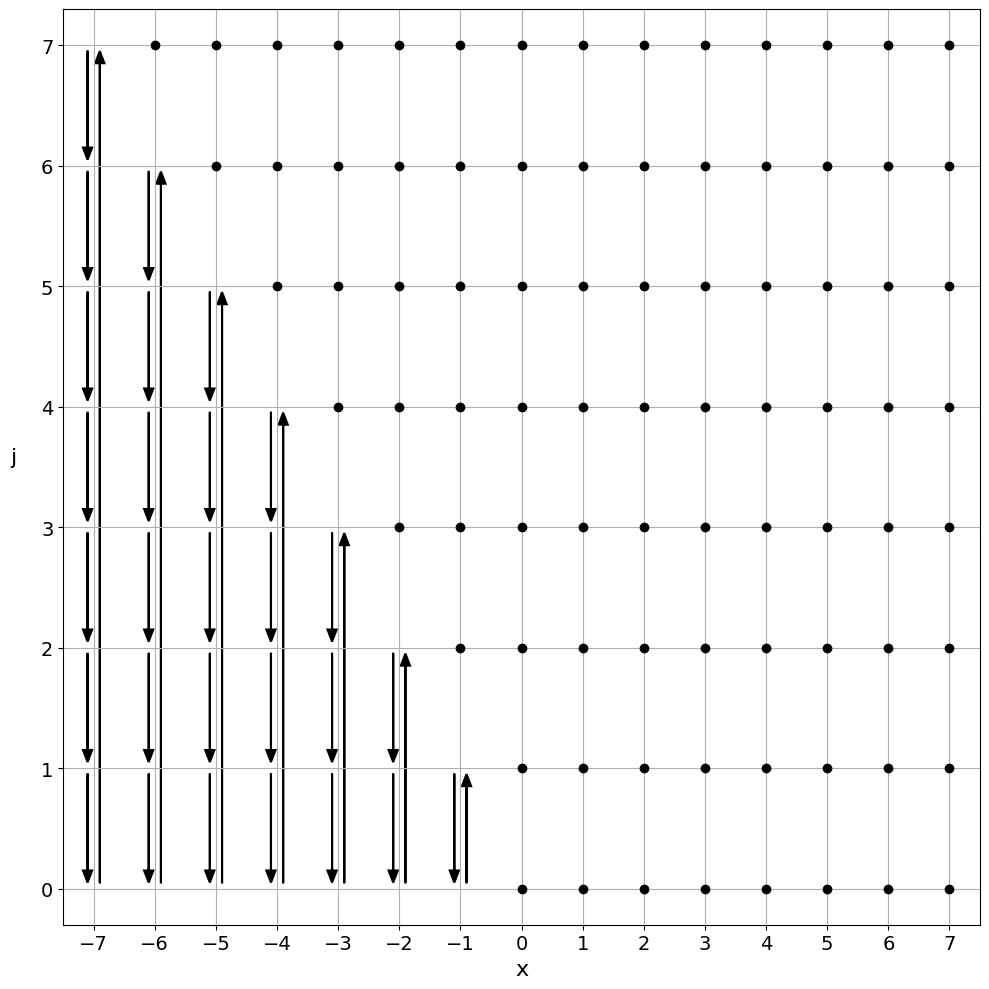}
\caption{Action of $W_+(U_1,U_0)$}
\label{fig_W+10}
\end{minipage}%
\begin{minipage}[b]{0.5\textwidth}
\centering
\includegraphics[width=0.7\textwidth]{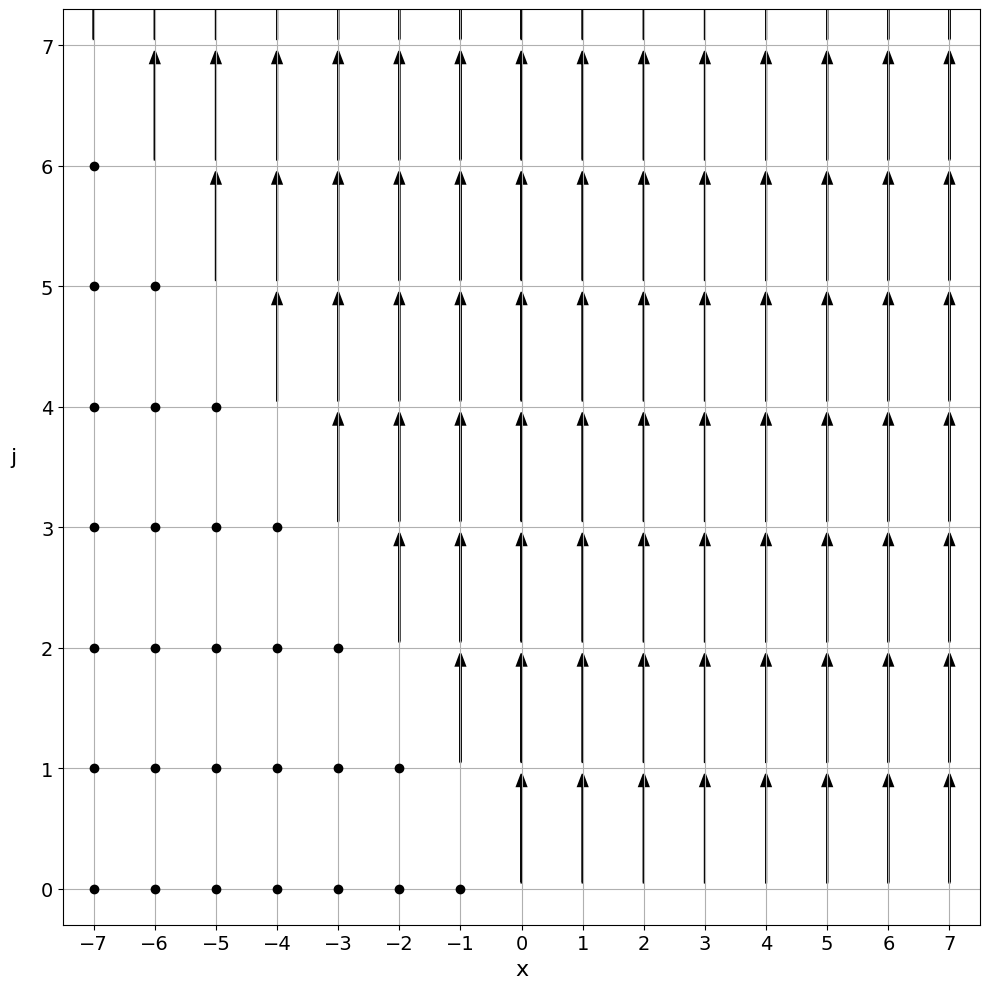}
\caption{Action of $W_-(U_1,U_0)$}
\label{fig_W-10}
\end{minipage}%
\end{figure}

In scattering theory, two of the main objects of interest are the wave operators, defined by
\begin{equation*}
W_\pm(U_1,U_0) :=s-\lim_{n\to \pm \infty} U_1^{-n}U_0^n,
\end{equation*}
where the limit is taken in the strong topology, which means when applied to an element of the Hilbert space.
By linearity, it is enough to consider the limits $\lim_{n\to \pm \infty} U_1^{-n}U_0^n \delta_{x,j}$,
which can be computed explicitly by using Figures \ref{fig_U_0} and \ref{fig_U_1}.
One then gets:
$$
W_+(U_1,U_0)\delta_{x,j}:=\begin{cases}
\delta_{x,j} & \hbox{ if }\  x\geq  -j+1, \\
\delta_{x,j-1} & \hbox{ if }\  x \leq  -j, j\geq 1, \\
\delta_{x,-x} & \hbox{ if }\  x \leq 0 , j= 0,
\end{cases}
$$
and
$$
W_-(U_1,U_0)\delta_{x,j}:=\begin{cases}
\delta_{x,j+1} & \hbox{ if }\  x \geq  -j, \\
\delta_{x,j} & \hbox{ if }\  x \leq -j-1.
\end{cases}
$$
We refer to Figures \ref{fig_W+10} and \ref{fig_W-10} for the
actions of these operators.

By looking carefully at the figures representing these two operators, one deduces their range, namely
$$
\Ran\big(W_\pm(U_1,U_0)\big)=\big\{W_\pm(U_1,U_0) \delta_{x,j}\mid x\in \Z,j\in \N\big\}.
$$
By inspection (see Remark \ref{rem_2}) one infers that $\Ran\big(W_+(U_1,U_0)\big)=\H$ 
and also that $\Ran\big(W_-(U_1,U_0)\big)\neq \H$. 
The orthocomplement of the range is called the cokernel. Looking more closely at these subspaces 
one obtains $\Coker\big(W_+(U_1,U_0)\big)=\{0\}$ and 
$\Coker\big(W_-(U_1,U_0)\big)=\Omega$, with $\Omega$ given by
\begin{equation}\label{eq_Omega}
\Omega:=\span\Big\{\delta_{x,j}\mid -x=j \in \N \hbox{ or } (x,j)\in \N\times \{0\}\Big\}\subset \H.
\end{equation}
Here, $\span$ means the subspace of $\H$ generated by the mentioned elements. 

\begin{rem}\label{rem_2}
On a figure representing an operator $A$, 
the range of the operator $A$ can be identified with the ending points of all arrows together with
the black dots. The cokernel of $A$ can be identified with the points which are neither the ending point
of any arrow nor a black dot.
\end{rem}

\begin{figure}[ht]
    \centering
    \includegraphics[width=5.5cm]{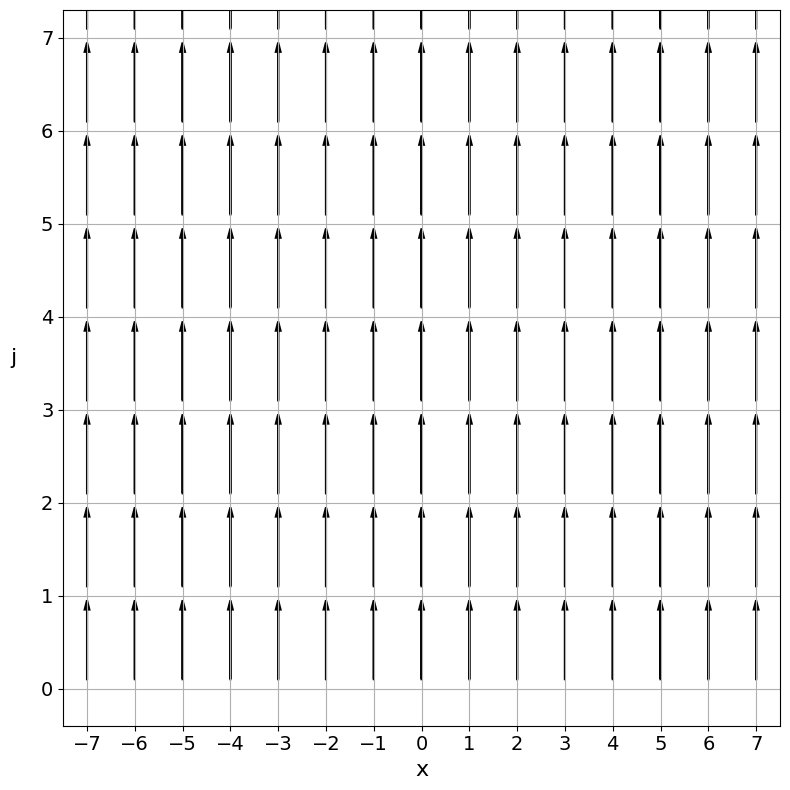}
    \caption{Action of $S(U_1,U_0)$}
   \label{fig_S10}
\end{figure}

In order to compute the scattering operator, we determine the adjoint of $W_+(U_1,U_0)$. 
From the equality 
$$
 \langle \delta_{y,k},W_+(U_1,U_0)^*\delta_{x,j}\rangle = 
\langle W_+(U_1,U_0) \delta_{y,k},\delta_{x,j}\rangle,
$$ 
one infers that
$$
W_+(U_1,U_0)^*\delta_{x,j}=\begin{cases}
\delta_{x,j} & \hbox{ if }\  x\geq  -j+1, \\
\delta_{x,j+1} & \hbox{ if }\  x \leq  -j-1,  \\
\delta_{x,0} & \hbox{ if }\  x =-j.
\end{cases}
$$
Then, the scattering operator $S(U_1,U_0):=W_+(U_1,U_0)^*W_-(U_1,U_0)$ acts on the orthonormal basis as 
$$
S(U_1,U_0)\delta_{x,j} =\delta_{x,j+1},
$$
where we refer to Remark \ref{rem_3} for an easy way to compute the product of two operators.
As a consequence, one easily observes that this operator is not unitary and that
$$
\Coker\big(S(U_1,U_0)\big)=\span\{\delta_{x,0}\mid x\in \Z\}.
$$
As for the wave operators, the scattering operator is only an isometry
(this fact is discussed in Section \ref{sec_outcomes}).
The action of $S(U_1,U_0)$ is represented in Figure \ref{fig_S10}.

\begin{rem}\label{rem_3}
If $A$ and $B$ correspond to operators represented in two figures, the action of the product
$BA$ on the element $\delta_{x,j}$ can be obtained by looking at the action of $B$ on $\delta_{y,k}$
with $\delta_{y,k}$ obtained by the relation $\delta_{y,k}=A\delta_{x,j}$. The resulting
element $\delta_{z,\ell} =B\delta_{y,k}$ represents the image of $\delta_{x,j}$ by the product $BA$.
\end{rem}

\subsection{Perturbed model 1}\label{sec1}

We shall now slightly perturb the model introduced in the previous section. 
For that purpose, we use the bra-ket notation, namely for any $f,g,h\in \H$
we set 
$$
|f\rangle \langle g| h:= \langle g,h\rangle f.
$$
The perturbation will be located around the position $(z,\ell)$ in the configuration space, 
on the right of the diagonal $\{(-j,j)\mid j\in \N\}$.

\begin{figure}[ht]
\centering
\begin{minipage}[b]{0.5\textwidth}
\centering
\includegraphics[width=0.7\textwidth]{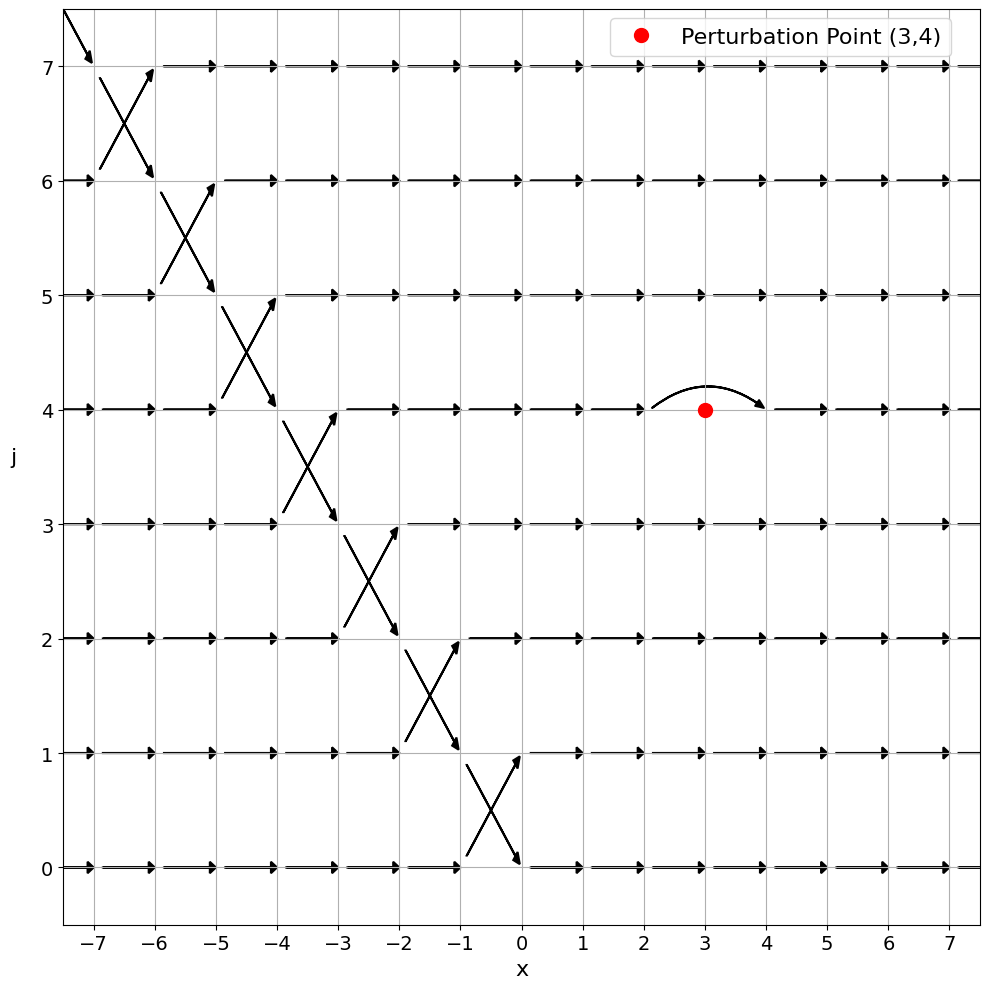}
\caption{Action of $U_2$}
\label{fig_U_2}
\end{minipage}%
\begin{minipage}[b]{0.5\textwidth}
\centering
\includegraphics[width=0.7\textwidth]{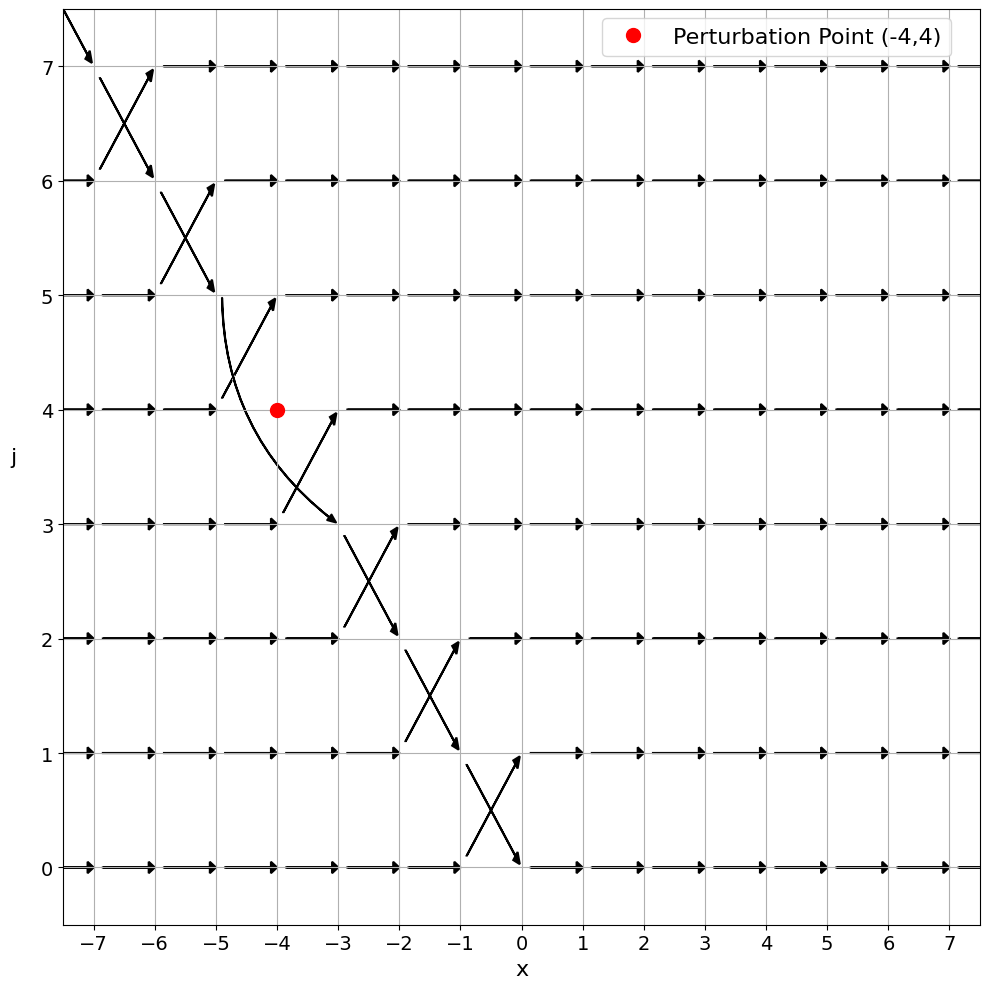}
\caption{Action of $U_3$}
\label{fig_U_3}
\end{minipage}%
\end{figure}

We fix $(z,\ell)\in \Z\times \N$ with $\ell\neq 0$ and $z\geq -\ell+2$. The new unitary operator
$U_2$ is defined by 
$$
U_2:=U_1-|\delta_{z,\ell}\rangle \langle \delta_{z-1,\ell}|-|\delta_{z+1,\ell}\rangle \langle \delta_{z,\ell}|
+ |\delta_{z+1,\ell}\rangle \langle \delta_{z-1,\ell}|+|\delta_{z,\ell}\rangle \langle \delta_{z,\ell}|.
$$
We refer to Figure \ref{fig_U_2} for the
action of this operator.

\begin{figure}[ht]
\centering
\begin{minipage}[b]{0.5\textwidth}
\centering
\includegraphics[width=0.7\textwidth]{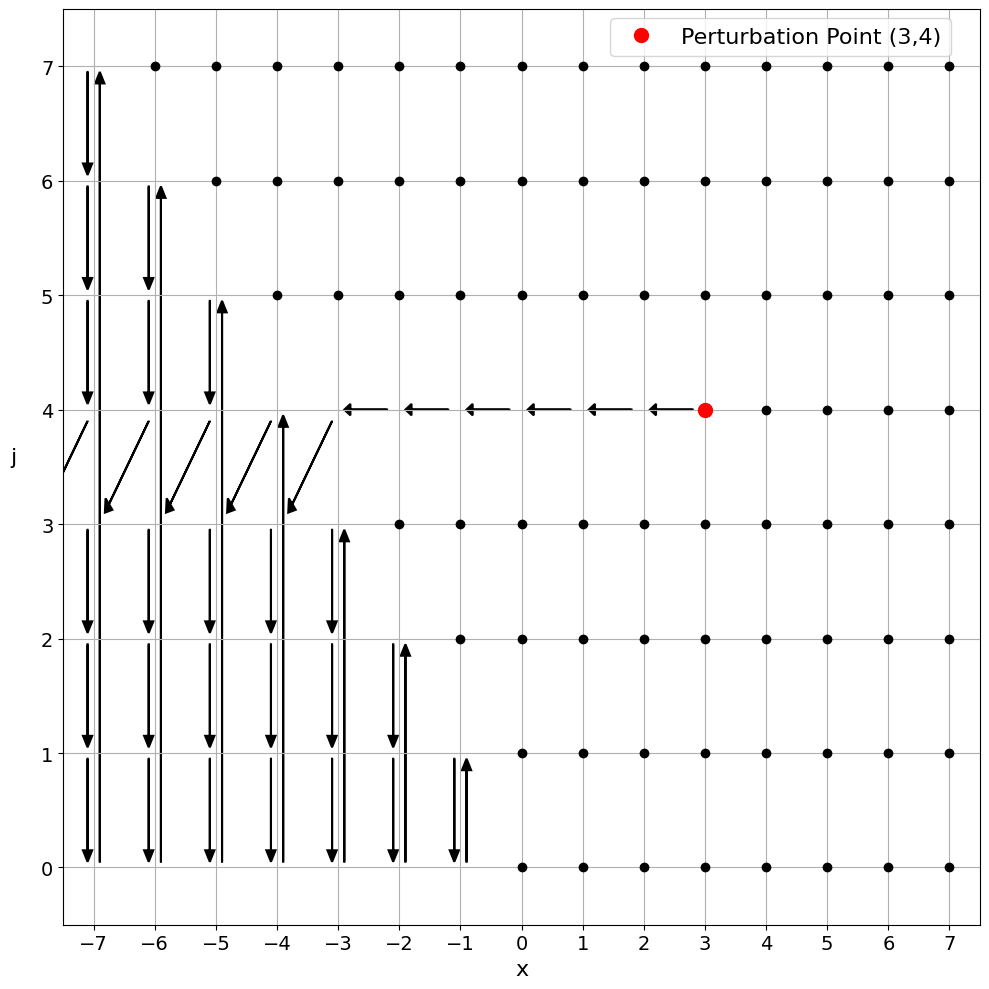}
\caption{Action of $W_+(U_2,U_0)$}
\label{fig_W+20}
\end{minipage}%
\begin{minipage}[b]{0.5\textwidth}
\centering
\includegraphics[width=0.7\textwidth]{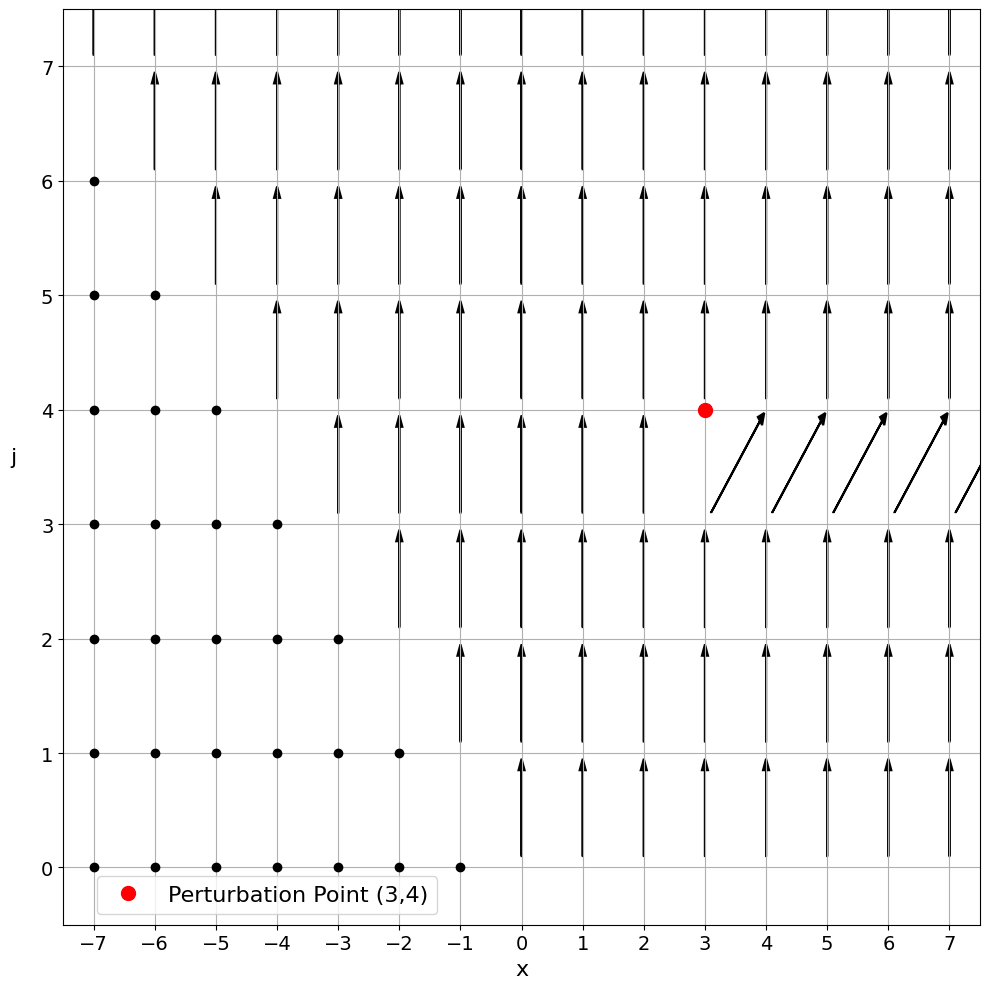}
\caption{Action of $W_-(U_2,U_0)$}
\label{fig_W-20}
\end{minipage}%
\end{figure}

By a direct computation, one gets 
$$
W_+(U_2,U_0)\delta_{x,j}:=\begin{cases}
\delta_{x,j} & \hbox{ if }\  j\neq \ell, x\geq  -j+1, \\
\delta_{x,j-1} & \hbox{ if }\  j\neq \ell, x \leq  -j, j\geq 1, \\
\delta_{x,-x} & \hbox{ if }\  j= 0, x \leq 0, \\ 
\delta_{x,\ell} & \hbox{ if }\  j=\ell, x\geq  z+1, \\
\delta_{x-1,\ell} & \hbox{ if }\  j=\ell, -\ell+2\leq x \leq  z, \\
\delta_{x-1,\ell-1} & \hbox{ if }\  j=\ell, x \leq -\ell+1,
\end{cases}
$$
and
$$
W_-(U_2,U_0)\delta_{x,j}:=\begin{cases}
\delta_{x,j} & \hbox{ if }\  x \leq -j-1, \\
\delta_{x,j+1} & \hbox{ if }\ j\neq \ell-1, x \geq  -j, \\
\delta_{x,j+1} & \hbox{ if }\  j=\ell-1, -\ell+1\leq x \leq z-1, \\
\delta_{x+1,j+1} & \hbox{ if }\  j=\ell-1, x \geq  z.
\end{cases}
$$
We refer to Figures \ref{fig_W+20} and \ref{fig_W-20} for the
actions of these operators.

From these results, one observes that the cokernels of these two operators are given by 
$\Coker\big(W_+(U_2,U_0)\big)=\C\delta_{z,\ell}$ and 
$\Coker\big(W_-(U_2,U_0)\big)=\span\{\delta, \delta_{z,\ell}\mid \delta\in \Omega\}$, respectively.
For the adjoint of $W_+(U_2,U_0)$ one gets
$$
W_+(U_2,U_0)^*\delta_{x,j}:=\begin{cases}
\delta_{x,j} & \hbox{ if }\  j\neq \ell, x\geq  -j+1, \\
\delta_{x,j+1} & \hbox{ if }\  j\neq \ell-1, x \leq  -j-1,  \\
\delta_{x,0} & \hbox{ if }\  j= -x,  \\ 
\delta_{x,\ell} & \hbox{ if }\  j=\ell, x\geq  z+1, \\
\delta_{x+1,\ell} & \hbox{ if }\  j=\ell, -\ell+1\leq x \leq  z-1, \\
\delta_{x+1,\ell} & \hbox{ if }\  j=\ell-1, x \leq -\ell,\\
0 & \hbox{ if } j=\ell, x=z.
\end{cases}
$$

\begin{figure}[ht]
\centering
\begin{minipage}[b]{0.5\textwidth}
\centering
\includegraphics[width=0.7\textwidth]{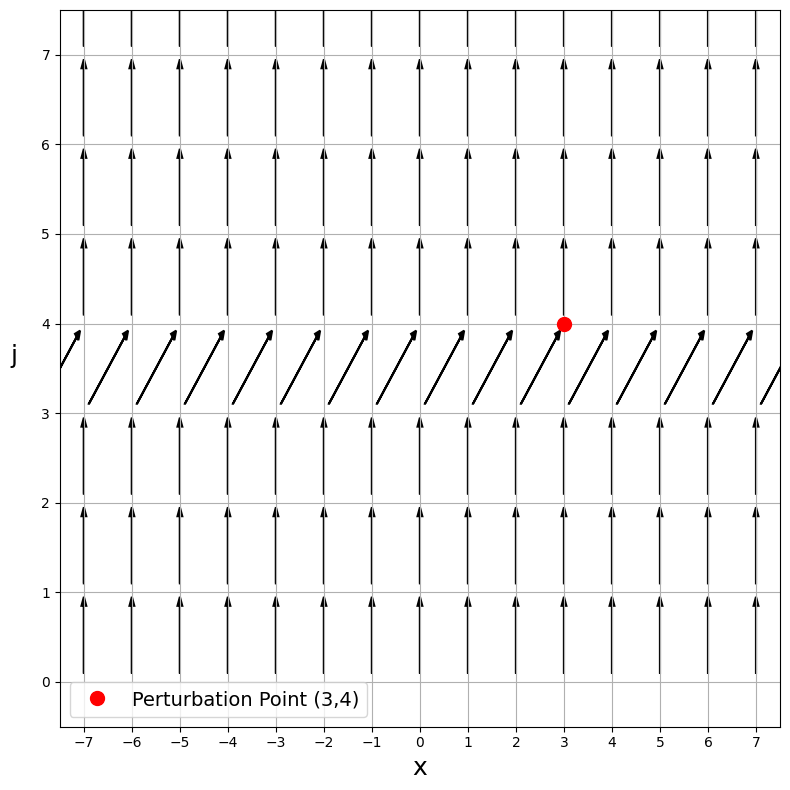}
\caption{Action of $S(U_2,U_0)$}
\label{fig_S20}
\end{minipage}%
\begin{minipage}[b]{0.5\textwidth}
\centering
\includegraphics[width=0.7\textwidth]{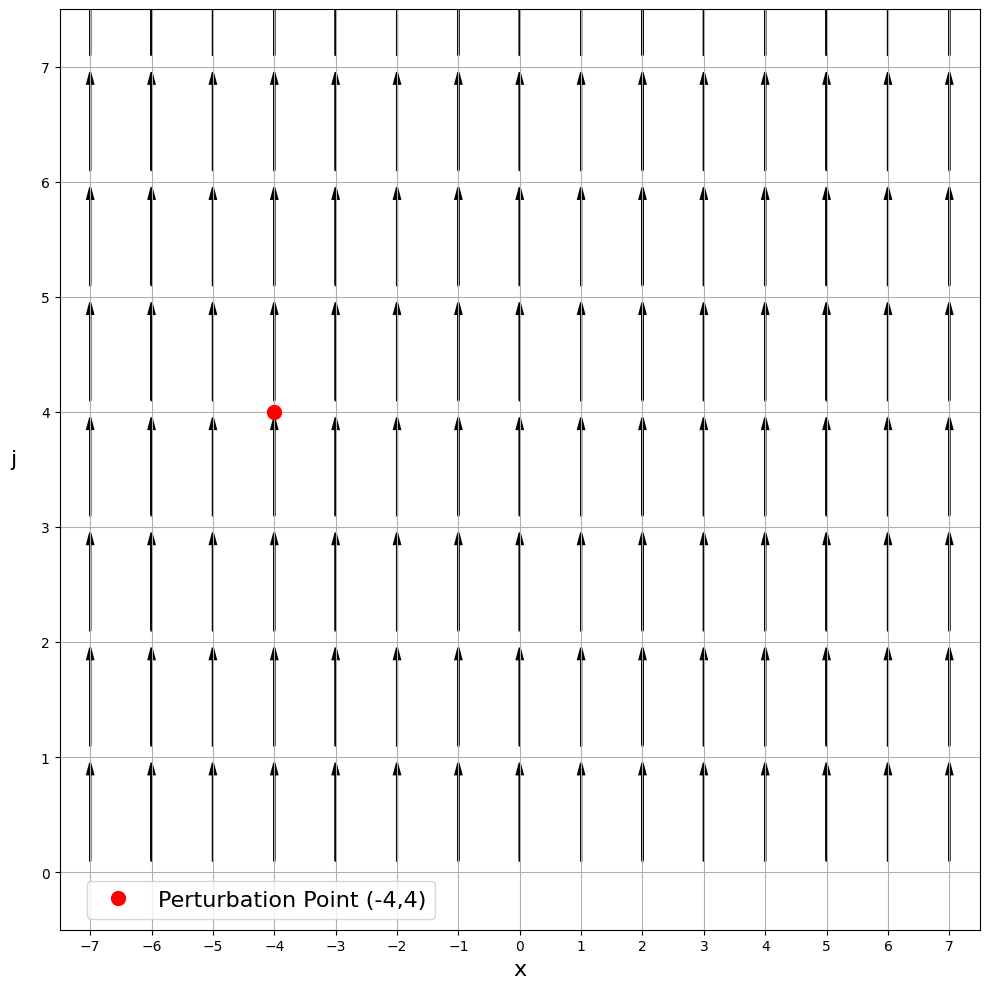}
\caption{Action of $S(U_3,U_0)$}
\label{fig_S30}
\end{minipage}%
\end{figure}

Then, the scattering operator $S(U_2,U_0):=W_+(U_2,U_0)^*W_-(U_2,U_0)$ is given by
$$
S(U_2,U_0)\delta_{x,j} =\begin{cases}
\delta_{x,j+1} & \hbox{ if }\  j\neq \ell-1, \\ 
\delta_{x+1,j+1} & \hbox{ if }\ j=\ell-1.
\end{cases}
$$
As before, one observes that 
$$
\Coker\big(S(U_2,U_0)\big)=\span\{\delta_{x,0}\mid x\in \Z\}.
$$
The action of $S(U_2,U_0)$ is represented in Figure \ref{fig_S20}.

\subsection{Perturbed model 2}\label{sec2}

We shall again perturb the initial model, but this time the perturbation
is located on the diagonal. More precisely, 
we fix $\ell\in \N$ with $\ell\geq 1$ and define the new unitary operator
$U_3$ by 
\begin{equation*}
U_3:= U_1-|\delta_{-\ell,\ell}\rangle \langle \delta_{-\ell-1,\ell+1}|-|\delta_{-\ell+1,\ell-1}\rangle \langle \delta_{-\ell,\ell}| 
+ |\delta_{-\ell+1,\ell-1}\rangle \langle \delta_{-\ell-1,\ell+1}|+|\delta_{-\ell,\ell}\rangle \langle \delta_{-\ell,\ell}|.
\end{equation*}
We refer to Figure \ref{fig_U_3} for the
action of this operator.

\begin{figure}[ht]
\centering
\begin{minipage}[b]{0.5\textwidth}
\centering
\includegraphics[width=0.7\textwidth]{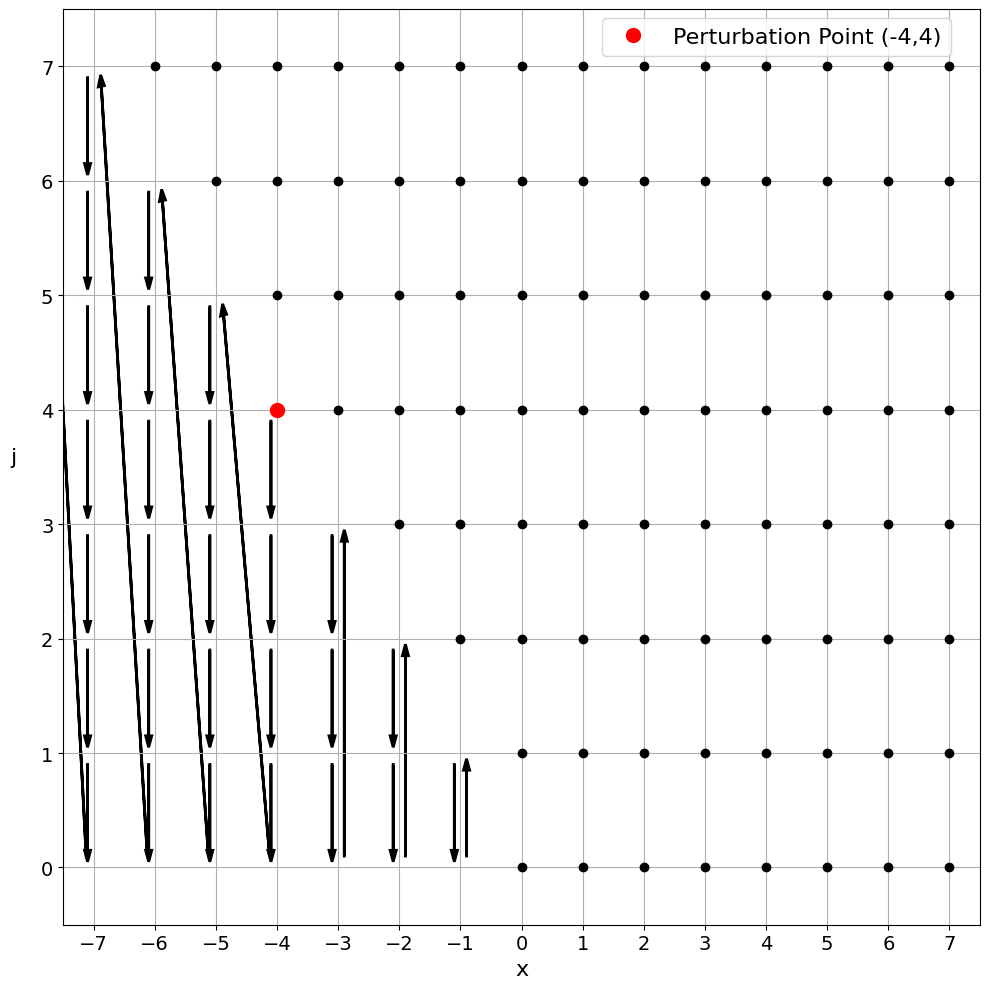}
\caption{Action of $W_+(U_3,U_0)$}
\label{fig_W+30}
\end{minipage}%
\begin{minipage}[b]{0.5\textwidth}
\centering
\includegraphics[width=0.7\textwidth]{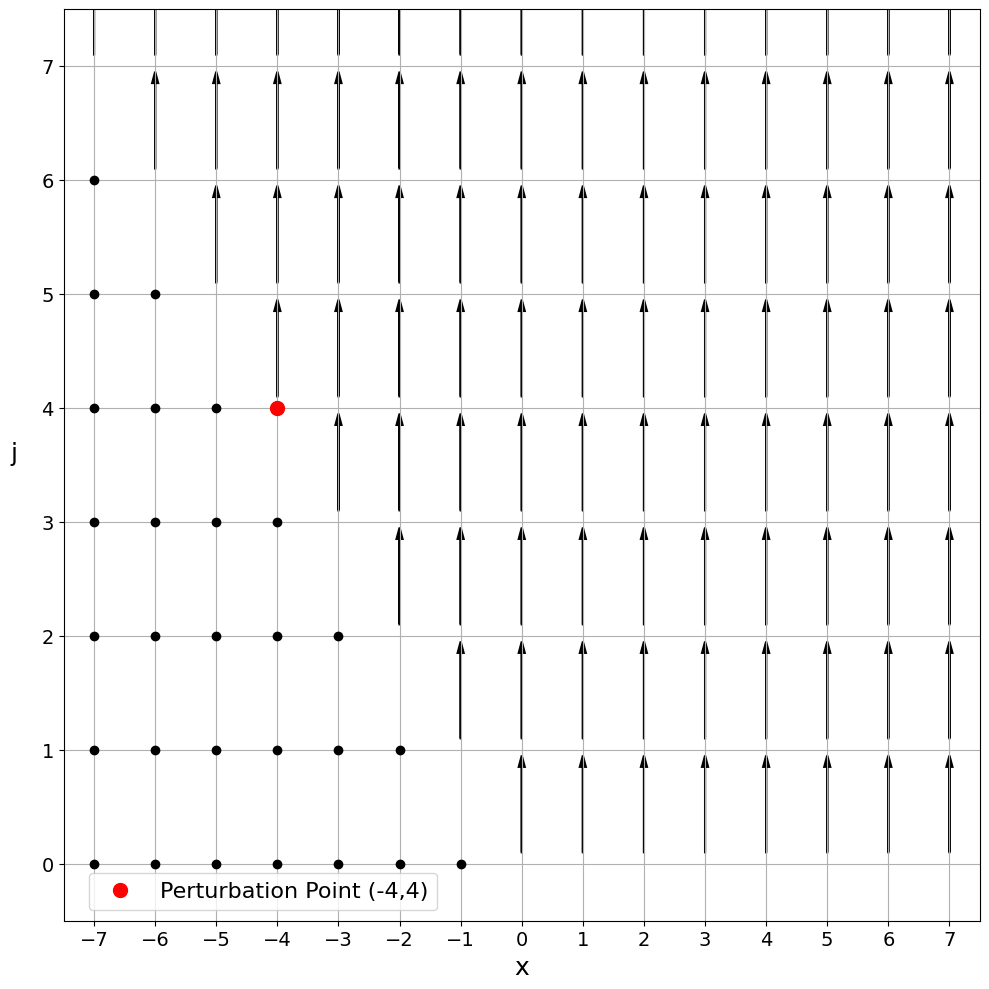}
\caption{Action of $W_-(U_3,U_0)$}
\label{fig_W-30}
\end{minipage}%
\end{figure}

By a direct computation, one gets 
$$
W_+(U_3,U_0)\delta_{x,j}:=\begin{cases}
\delta_{x,j} & \hbox{ if }\  x\geq  -j+1, \\
\delta_{x,j-1} & \hbox{ if }\  j\neq 0, x \leq  -j, \\
\delta_{x,-x} & \hbox{ if }\  j= 0, -\ell+1 \leq x \leq 0,  \\ 
\delta_{x-1,-x+1} & \hbox{ if }\  j=0, x \leq -\ell,
\end{cases}
$$
and
$$
W_-(U_3,U_0)\delta_{x,j}:=\begin{cases}
\delta_{x,j} & \hbox{ if }\  x \leq -j-1, \\
\delta_{x,j+1} & \hbox{ if }\ x\geq -j.
\end{cases}
$$
We refer to Figures \ref{fig_W+30} and \ref{fig_W-30} for the
actions of these operators.

This gives $\Coker\big(W_+(U_3,U_0)\big)=\C\delta_{-\ell,\ell}$ and $\Coker\big(W_-(U_3,U_0)\big)=\Omega$.
For the adjoint of $W_+(U_3,U_0)$ one gets
$$
W_+(U_3,U_0)^*\delta_{x,j}:=\begin{cases}
\delta_{x,j} & \hbox{ if }\  x\geq  -j+1, \\
\delta_{x,j+1} & \hbox{ if }\  x \leq  -j-1, \\
\delta_{x,0} & \hbox{ if }\  j= -x, -\ell+1 \leq x \leq 0,  \\ 
\delta_{x+1,0} & \hbox{ if }\  j=-x, x \leq -\ell-1, \\
0 & \hbox{ if } j=\ell, x=-\ell.
\end{cases}
$$
Then, the scattering operator $S(U_3,U_0):=W_+(U_3,U_0)^*W_-(U_3,U_0)$ is given by
$$
S(U_3,U_0)\delta_{x,j} =\delta_{x,j+1},
$$
which has an action similar to the one of $S(U_1,U_0)$, see Figure \ref{fig_S30} and the comment in the next section.

\section{Scattering outcomes}\label{sec_outcomes}

In this section, we start by recalling a few  general properties of scattering systems, and then 
illustrate these properties on the models introduced in the previous section.

\subsection{General properties}

Let us assume that the following wave operators exist:
\begin{equation}\label{eq_Wpm_2}
W_\pm(U,U_0) :=s-\lim_{n\to \pm \infty} U^{-n}U_0^n,
\end{equation}
where $U,U_0$ are unitary operators on a Hilbert space $\H$.
Most of the subsequent properties are based on the intertwining relation. 
More precisely, for any $m\in \Z$ one has
$$
W_\pm(U,U_0)\;\! U_0^m = U^m \;\!W_\pm(U,U_0)
$$
These general relations can be directly deduced
from \eqref{eq_Wpm_2}, by a change of variable in the limits.

From these relations, one firstly deduces the invariance of the range of the wave operators, 
and accordingly of the cokernel of these operators. 
Indeed, since by definition $\Ran\big(W_\pm(U,U_0)\big) = W_\pm (U,U_0)\H$ one has
\begin{equation*}
U W_\pm (U,U_0)\H = W_\pm(U,U_0) U_0 \H  = W_\pm(U,U_0) \H.
\end{equation*}
where we have used that $U_0\H=\H$. These equalities mean that 
$$
U\;\!\Ran\big(W_\pm(U,U_0)\big)=\Ran\big(W_\pm(U,U_0)\big).
$$
Since $\Coker\big(W_\pm(U,U_0)\big) = \big(\Ran\big(W_\pm(U,U_0)\big)\big)^\bot$, one readily deduces
that
$$
U\;\!\Coker\big(W_\pm(U,U_0)\big)=\Coker\big(W_\pm(U,U_0)\big).
$$

Another general property which follows from the intertwining relation is the commutations relation between the scattering
operator $S(U,U_0)$ and the free evolution $U_0$. More precisely one has
\begin{equation*}
U_0\;\! W_+(U,U_0)^*\;\! W_-(U,U_0) = W_+(U,U_0)^* \;\!U\;\! W_-(U,U_0)  = W_+(U,U_0)^*\;\! W_-(U,U_0) \;\!U_0 
\end{equation*}
leading to the equality $U_0\;\! S(U,U_0) = S(U,U_0)\;\!U_0$.
As illustrated soon on the models, the commutation relation $[S(U,U_0),U_0]=0$ implies a very strong constraint on the form of the scattering operator.

We finally mention one more property related to the wave operators: the chain rule. It says that if one considers three unitary
operators and if the wave operators exist for two pairs, then they also exist for the third pair, provided a certain compatibility
condition (ordering) is satisfied. We shall illustrate this rule explicitly on our examples.

\subsection{Illustration on the models}

We now come back to the models of Section \ref{sec:models} and illustrate the previous relations on them.

\subsubsection{Invariance of cokernels}

Firstly, the invariance of the cokernels can be checked by hand. Indeed, one observes that
$U_1\Omega=\Omega$, that $U_2\Omega=\Omega$, and that $U_2 \delta_{z,\ell} = \delta_{z,\ell}$.
Similarly, one has $U_3\Omega=\Omega$, and $U_3 \delta_{-\ell,\ell} = \delta_{-\ell,\ell}$. 
In addition, the following invariance also holds: $U_3\Omega'=\Omega'$ with 
\begin{equation}\label{eq_Omega'}
\Omega':=\span\Big\{\delta_{x,j}\mid -x=j \in \N\setminus \{\ell\} \hbox{ or } (x,j)\in \N\times \{0\}\Big\}\subset \H.
\end{equation}

As explained in the Introduction, elements in the cokernel of the wave operators can not be described
asymptotically by the evolution group generated by $U_0$. By looking at $\Coker\big(W_\pm(U_i,U_0)\big)$
for $i\in \{1,2,3\}$, this fact becomes rather clear. Indeed the action of $U_1$ on $\Omega$, of $U_2$ on $\Omega$,
and of $U_3$ on $\Omega'$ are equivalent to the action of a shift operator on $\ell^2(\Z)$, but
the iterated actions $U_1^{n}$, $U_2^{n}$ and $U_3^{n}$ on these respective domains have 
nothing to do with $U_0^{n}$ when $n\to -\infty$, see Figures \ref{fig_U_0}, \ref{fig_U_1}, \ref{fig_U_2}, and \ref{fig_U_3}.
This explains why the cokernel of $W_-(U_i,U_0)$ is not empty, for $i\in \{1,2,3\}$.
On the other hand, when $n\to \infty$, the asymptotic evolution generated by $U_0^n$ matches the asymptotic
evolution generated by $U_i^n$, assuring the existence of the wave operators $W_+(U_i,U_0)$ with 
nothing in the cokernel except the bound states of $U_i$.
As a result of this asymmetry in the ranges of $W_+(U_i,U_0)$ and $W_-(U_i,U_0)$,
the rather common equality $\Ran\big(W_+(U_i,U_0)\big) = \Ran\big(W_-(U_i,U_0)\big)$
does not hold for our examples. The main consequence is that the scattering operator $S(U_i,U_0)$
is not unitary, while this property holds when the ranges are equal.

Let us still observe that the $1$-dimensional subspace spanned by $\delta_{z,\ell}$ is left invariant by
the action on $U_2$, and the $1$-dimensional subspace spanned by $\delta_{-\ell,\ell}$ is left invariant by $U_3$.
These subspaces correspond to bound states of the operators $U_2$ and $U_3$, respectively. 
Clearly, the evolution on these subspaces has nothing to do with the evolution generated by $U_0$,
and therefore they belong to the cokernels of $W_\pm(U_2,U_0)$ and $W_\pm(U_3,U_0)$, respectively.

\subsubsection{Commutation between scattering operators and free evolution}

Let us now illustrate the commutation relation of the scattering operators with the free evolution $U_0$.
This commutation relation is clearly visible in the figures. Indeed, the action of 
$S(U_i,U_0)$ on $\delta_{x,j}$ is independent of $x\in \Z$, for $i\in \{1,2,3\}$, 
see Figures \ref{fig_S10}, \ref{fig_S20}, and \ref{fig_S30}. 
Since the action of $U_0$ is precisely a shift on the $x$-variable, the commutation relation holds automatically.
In fact, any operator commuting with $U_0$ must have an action on $\delta_{x,j}$ which is independent of $x$.

As already observed for the perturbed model 2, the equality $S(U_3,U_0)=S(U_1,U_0)$ holds, even if $U_3$ and $U_1$
are different. Clearly, this equality has a rather deep consequence: by just knowing the scattering operator, 
it is not possible to make any difference between the initial scattering system modeled by the evolution $U_1$ and
the perturbed evolution driven by $U_3$. Thus, the uniqueness of the inverse scattering problem based only on the 
scattering operator will not be possible for these systems. However, the prior knowledge of the existence of a bound state from $U_3$ and
the absence of any bound state for $U_1$ might help for the inverse problem. We do not investigate the inverse
problem any further here.

Let us still add one remark on the scattering operators. Clearly, the operators $U_1$ and $U_2$ differ
only by a finite rank perturbation. However, the operators $S(U_1,U_0)$ and $S(U_2,U_0)$ are quite different,
and this difference is not even compact. This feature is rather well-known in scattering theory: 
a small change in the initial evolution operators can lead to a drastic change in the resulting
scattering operators. The perturbed model 1 is a good illustration of this phenomenon.

\subsubsection{Chain rule}\label{sec_cr}

\begin{figure}[ht]
\centering
\begin{minipage}[b]{0.5\textwidth}
\centering
\includegraphics[width=0.7\textwidth]{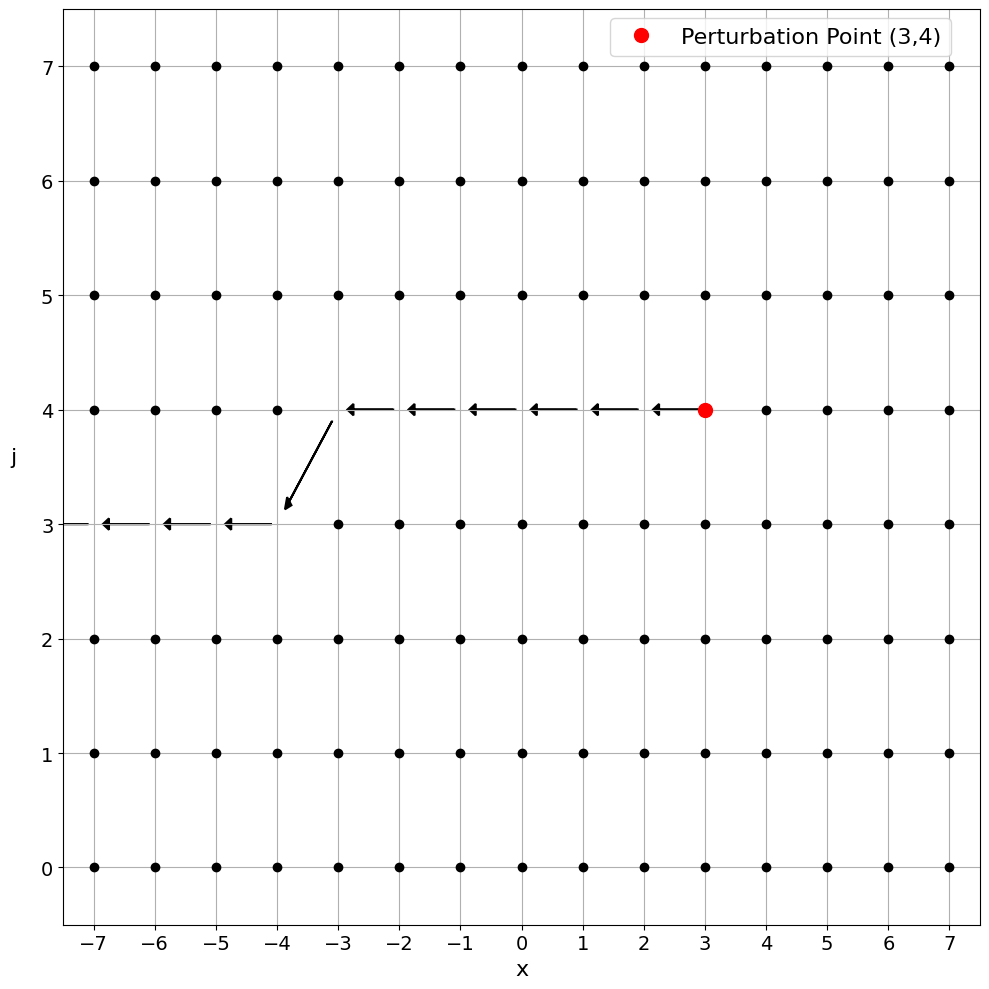}
\caption{Action of $W_+(U_2,U_1)$}
\label{fig_W+21}
\end{minipage}%
\begin{minipage}[b]{0.5\textwidth}
\centering
\includegraphics[width=0.7\textwidth]{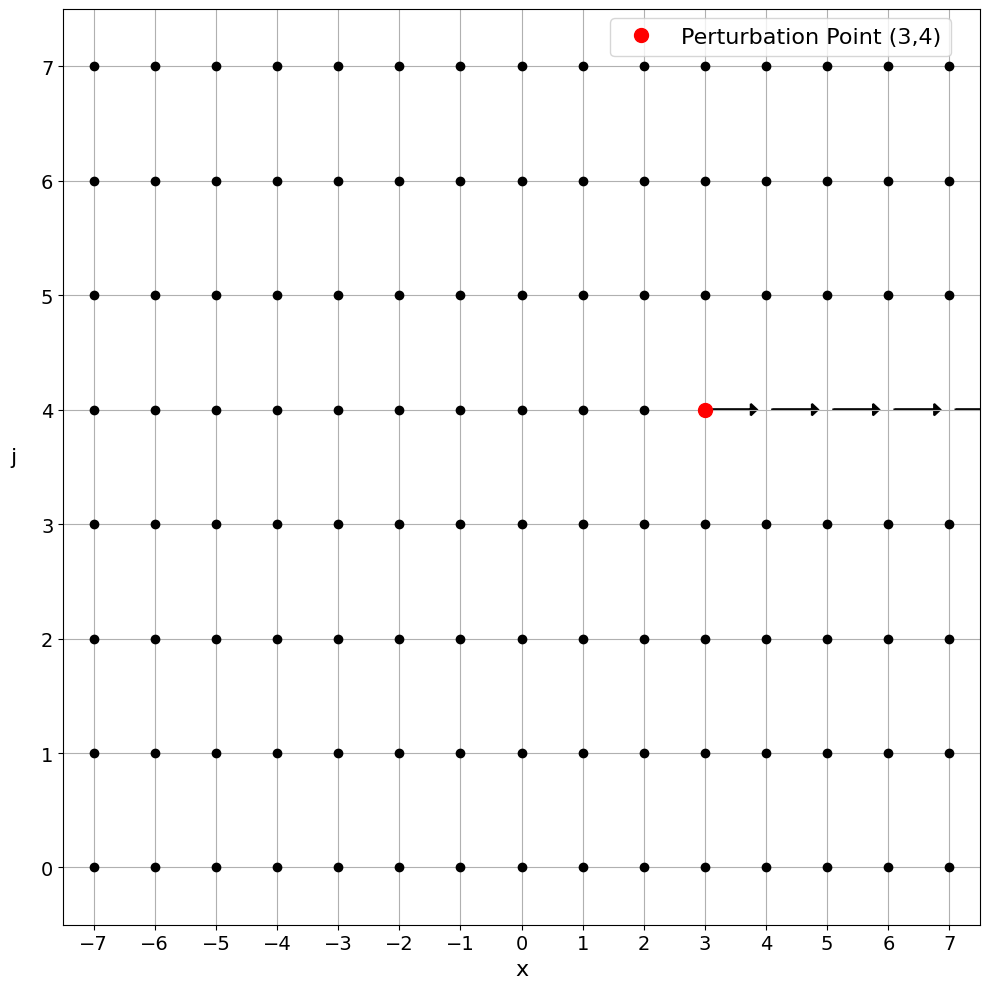}
\caption{Action of $W_-(U_2,U_1)$}
\label{fig_W-21}
\end{minipage}%
\end{figure}

The wave operators $W_\pm(U_2,U_0)$ have already been computed in the previous section, but let us show that
they can be obtained by a suitable product. For that purpose, one needs to compute
$W_\pm (U_2,U_1)$. This can be readily done and one obtains
$$
W_+(U_2,U_1)\delta_{x,j}:=\begin{cases}
\delta_{x,j} & \hbox{ if }\  j\neq \ell, x\geq  -j, \\
\delta_{x,j} & \hbox{ if }\  j\neq \ell-1, x\leq  -j, \\
\delta_{x,\ell} & \hbox{ if }\  j=\ell, x\geq  z+1, \\
\delta_{x-1,\ell} & \hbox{ if }\  j=\ell, -\ell+2\leq x \leq  z, \\
\delta_{x-1,\ell-1} & \hbox{ if }\  j=\ell, x=-\ell+1, \\
\delta_{x-1,\ell-1} & \hbox{ if }\  j=\ell-1, x \leq -\ell.
\end{cases}
$$
$$
W_-(U_2,U_1)\delta_{x,j}:=\begin{cases}
\delta_{x,j} & \hbox{ if }\  j\neq \ell, \\
\delta_{x,\ell} & \hbox{ if }\  j=\ell, x\leq  z-1, \\
\delta_{x+1,\ell} & \hbox{ if }\  j=\ell,  x \geq  z.
\end{cases}
$$
We refer to Figures \ref{fig_W+21} and \ref{fig_W-21} for the
actions of these operators.

By considering then the following products one infers the equalities
$$
W_\pm(U_2,U_1) W_\pm(U_1,U_0) = W_\pm(U_2,U_0),
$$
where $W_\pm(U_2,U_0)$ were illustrated in Figure \ref{fig_W+20} and \ref{fig_W-20}.
These relations correspond to the chain rule and hold in a rather general framework.

\section{Wold's decomposition}\label{sec_Wold}

Let us first introduce one notation: we set $\Sh$ for the shift operator acting on $\ell^2(\N)$. 
On the standard basis $\{\delta_n\}_{n\in\N}$ of $\ell^2(\N)$, this operator acts as $\Sh \delta_n=\delta_{n+1}$.
We also recall that an isometry $W$ on a Hilbert space $\H$ satisfies $\|Wf\|=\|f\|$, or equivalently
$W^*W = 1$. In particular, $\Sh$ is an isometry. We now state Wold's decomposition:

\begin{thm}[Thm.~V.2.1 of \cite{Davidson}]
If $W$ is an isometry on a Hilbert space $\H$, then there is a cardinal number $\alpha$
and a unitary operator $U$ (possibly vacuous) such that $W$ is unitarily equivalent to $\Sh^{(\alpha)}\oplus U$.
\end{thm}

Let us show that all isometries introduced in Section \ref{sec:models} fit well with
this statement and the unitary equivalence can be implemented explicitly.
We start with the operators $W_\pm(U_1,U_0)$ of Section \ref{sec0}.
Recall that the operator $W_+(U_1,U_0)$ is unitary, while the operator $W_-(U_1,U_0)$ is only an isometry, 
with a cokernel described by \eqref{eq_Omega}. 
Obviously, there is nothing special to say in relation to Wold's decomposition for the operator $W_+(U_1,U_0)$.
Still, it is interesting to observe on Figure \ref{fig_W+10} that the unitary character of $W_+(U_1,U_0)$
is due to two distinct types of orbits: fixed points for $\delta_{x,n}$ for $x\geq -n+1$, and cyclic orbits
otherwise.

For $W_-(U_1,U_0)$ represented in Figure \ref{fig_W-10}, the key observation is that this isometry
has already a form that suits well with Wold's decomposition.
Indeed, let us provide an alternative form of the Hilbert space $\H$, namely
$$
\H= \H_{00} \oplus \bigoplus_{x\in \Z} \H_x
$$
with 
$$
\H_{00} =\span\Big\{\delta_{x,j} \mid (x,j)\in \Z\times \N \hbox{ with }x\leq -j-1\Big\}
$$
and for $x\in \Z$
$$
\H_x =\span\Big\{\delta_{x,j} \mid j\in \N \hbox{ with }j\geq \min\{-x,0\}\Big\}\cong \ell^2(\N)
$$
Then, $W_-(U_1,U_0)$ leaves each of these subspaces invariant, and $W_-(U_1,U_0)\big|_{\H_{00}}=1$
while $W_-(U_1,U_0)\big|_{\H_x}=\Sh$.
Thus, in relation to Wold's decomposition, the cardinal number $\alpha$ is $0$ for $W_+(U_1,U_0)$, while
this cardinal number is $\aleph_0$\footnote{Called ``aleph-zero". It represents the cardinality of the set $\Z$.} 
for $W_-(U_1,U_0)$.

The situation for $W_-(U_2,U_0)$, represented in Figure \ref{fig_W-20} is quite similar: 
A decomposition of the Hilbert space $\H_{00}$ and $\oplus_{x\in \Z}\H_x'$, with $\H_x'\cong \ell^2(\N)$
and the action of $W_-(U_2,U_0)$ on $\H_x'$ unitarily equivalent to the shift operator $\Sh$.
As a consequence $\alpha=\aleph_0$ for $W_-(U_2,U_0)$.

On the other hand, the operator $W_+(U_2,U_0)$ represented in Figure \ref{fig_W+20} shows a decomposition
of this operator into a unitary part made of fixed points and $(\ell-1)$ non-trivial cycles, and one
part (not easy to describe but easy to visualize) unitarily equivalent to the shift operator
$\Sh$ in $\ell^2(\N)$. For concreteness, let us mention that the point $0$ of $\ell^2(\N)$ coincides with
the point $(z,\ell)$ of the perturbed system. At the end of the day, $\alpha=1$ for the operator $W_+(U_2,U_0)$.

Finally, the isometry $W_-(U_3,U_0)$ represented in Figure \ref{fig_W-30} is quite similar to the operator $W_-(U_1,U_0)$
with respect to Wold's decomposition. On the other hand, the wave operator $W_+(U_3,U_0)$ represented in Figure \ref{fig_W+30}
is quite similar to $W_+(U_2,U_0)$. Indeed, this operator can be decomposed into a unitary part made
of fixed points and $\ell-1$ non-trivial cycles, and one part unitarily equivalent to a shift operator $\Sh$
in the Hilbert space $\ell^2(\Z)$. Again, this part is easy to visualize on Figure \ref{fig_W+30} but 
less easy to describe. 
Let us mention that the point $0$ of $\ell^2(\N)$ coincides with
the point $(-\ell,\ell)$ of the perturbed system, and that $\alpha=1$ for the operator $W_+(U_3,U_0)$.

Note that all scattering operators and the wave operators $W_\pm(U_2,U_1)$ of Section \ref{sec_cr}
are also isometries, and therefore the same observations with respect to Wold's decomposition hold also for them.
We leave this simple exercise for the interested reader.

\begin{rem}
Let us mention that Wold's decomposition has already been linked to scattering theory a long time ago, 
see for example \cite{Masani, SN}. In particular, it has been connected to the Lax-Phillips approach
of scattering theory. However, we could not find explicit descriptions or pictures as provided in the
present work.
\end{rem}

\section{Conclusion}
The model introduced in this paper allows us to compute explicitly all objects of the related discrete
time scattering theory. Various properties of these operators are clearly illustrated and discussed.
Dynamical systems with so explicit expressions are rather rare and should be used for introducing the theory
at a pedagogical level. Finally, the expressions obtained illustrate a rather deep theorem of operator theory,
Wold's decomposition of isometries.

\end{document}